\documentclass[aps,twocolumn,epsf,floats,pre]{revtex4-1}
\usepackage{graphics,graphicx,epsfig}
\usepackage{amssymb,color}
\usepackage{epsf,epstopdf,wrapfig}
\usepackage{amsmath}

\begin{document}

\title{Superfluid transport of information in turning flocks of starlings}

\author{Alessandro Attanasi$^{*,\ddagger}$, Andrea Cavagna$^{*,\ddagger}$, Lorenzo Del Castello $^{*,\ddagger}$, Irene Giardina$^{*,\ddagger}$, Tomas S. Grigera$^\flat$, Asja Jeli\'c$^{*,\ddagger}$, Stefania Melillo$^{*,\ddagger}$, Leonardo Parisi$^{*,\S}$, Oliver Pohl$^{*,\ddagger}$, Edward Shen$^{*,\ddagger}$, Massimiliano Viale$^{*,\ddagger}$}

\affiliation{$^*$ Istituto Sistemi Complessi, Consiglio Nazionale delle Ricerche, UOS Sapienza, 00185 Rome, Italy}

\affiliation{$^\ddagger$ Dipartimento di Fisica, Universit\`a\ Sapienza, 00185 Rome, Italy}

\affiliation{$^\flat$ Instituto de Investigaciones Fisicoqu{\'\i}micas
  Te{\'o}ricas y Aplicadas (INIFTA) and Departamento de F{\'\i}sica,
  Facultad de Ciencias Exactas, Universidad Nacional de La Plata,
  c.c. 16, suc. 4, 1900 La Plata, Argentina} \affiliation{CONICET La
  Plata, Consejo Nacional de Investigaciones Cient{\'\i}ficas y
  T{\'e}cnicas, Argentina}

\affiliation{$^\S$ Dipartimento di Informatica, Universit\`a\ Sapienza, 00198 Rome, Italy}

\begin{abstract}
Collective decision-making in biological systems requires all individuals in the group to go through 
a behavioural change of state. During this transition, the efficiency of information transport is a key factor
to prevent cohesion loss and preserve robustness. The precise mechanism by which natural groups
achieve such efficiency, though, is currently not fully understood.
Here, we present an experimental study of starling flocks performing collective turns in the field. 
We find that the information to change direction propagates across the flock linearly 
in time with negligible attenuation, hence keeping group decoherence to a minimum. 
This result contrasts with current theories of collective motion, which 
predict a slower and dissipative transport of directional information. 
We propose a novel theory whose cornerstone is the existence of a conserved spin current generated 
by the gauge symmetry of the system. The theory turns out to be mathematically identical 
to that of superfluid transport in liquid helium and it explains the dissipationless propagating 
mode observed in turning flocks. Superfluidity also provides a quantitative expression for the speed of 
propagation of the information, according to which transport must be swifter the stronger the group's
orientational order. This prediction is verified by the data. 
We argue that the link between strong order and efficient decision-making required by superfluidity 
may be the adaptive drive for the high degree of behavioural polarization observed in many living groups.
The mathematical equivalence between superfluid liquids and turning flocks is a compelling demonstration of the 
far-reaching consequences of symmetry and conservation laws across different natural systems.
\end{abstract}


\maketitle


Consider a flock of starlings under direct threat from a peregrine falcon. To dodge the attack, every split second the group collectively takes the decision to change direction of motion \cite{krause+ruxton_02}. Each such change, however, puts the flock in a vulnerable condition, which the predator is precisely there to exploit. The slightest uncertainty may decrease cohesion, or even split the group and push some birds astray, leaving them easy prey of the falcon, thus decreasing the fitness of the group.
This is a general issue for collective decision-making in social species, for which forming cohesive groups is a matter of fitness, either for anti-predatory reasons, or other environmental concerns \cite{camazine+al_01,conradt+Roper_05,conradt+List_09,conradt+al_09}. Irrespective of what is the consensus-forming mechanism leading to the decision, its actual execution cannot be instantaneous, as a certain time lag is needed to propagate the decision throughout the group. During this time there is a transient mixture of individuals who have already changed state and individuals  who have not yet done so.  For this reason a behavioural change of state is intrinsically in conflict with cohesion.  Consensus must be tight and the decision must spread across the group swiftly enough to guarantee robustness of the group \cite{couzin+krause_03,sumpter_08}.

A collective change of state may be the result of a perturbation hitting most individuals in the group. For example, a shot heard by an entire flock of birds sitting on a tree makes them all take off  at the same time. When this happens there is hardly any transport of information.
More interesting is the case when the collective decision has a localized spatial origin, starting from a few individuals close to each other. This may be due either to an external stimulus (a predator, for example) or to some spontaneous behavioural fluctuation. In this case, the information to change state must spread across the group and reach all individuals.  It becomes therefore essential to understand the mechanism of transport of the information. Does the signal get attenuated in space and time? What are the laws of propagation and how do they depend on the parameters of the group? These questions have a major impact on the functional effectiveness of collective decision-making. However, little is known about these problems, both from the empirical and the theoretical point of view \cite{parrish_97,bajec+heppner_09,nagy+al_10}.

 \begin{figure*}[t!] 
  \centering
  \includegraphics[width=1.8\columnwidth]{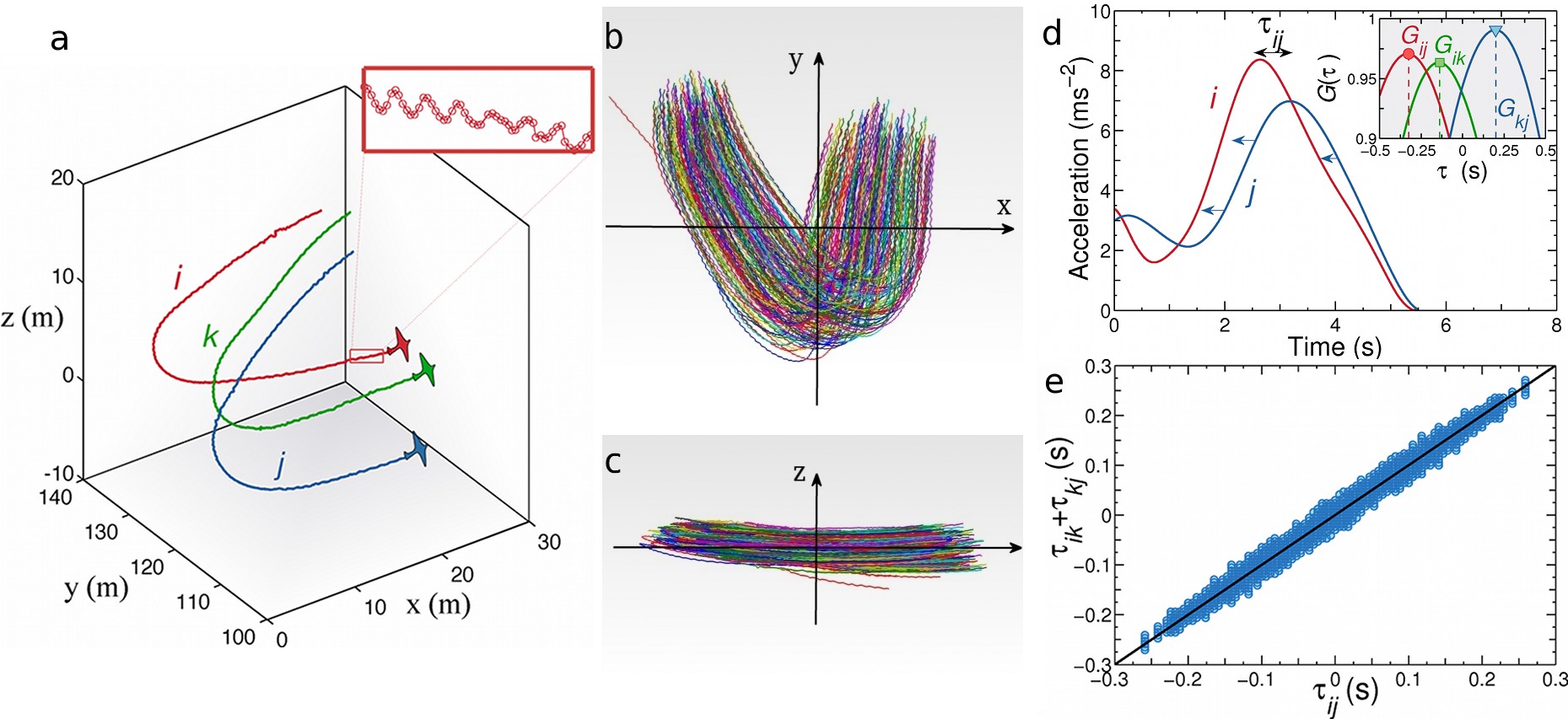}
 \caption{
{\bf Birds trajectories and turning delays.}
{\bf a,} 
Reconstructed $3d$ trajectories of three birds belonging to a flock performing a collective turn. Sampling at $170$Hz we capture fine details of the birds
movement, such as the zig-zag due to wing flapping ($10$Hz in starlings - inset). 
{\bf b, c,} 
Trajectories of all $N=176$ birds of the same flock as in panel {\bf a}. Each trajectory lies approximately on a plane, justifying a simplified planar description of the velocity.
{\bf d,} 
The radial acceleration of a turning bird displays a maximum as a function of time.
In principle, given two birds $i$ and $j$, one could simply define the turning delay $\tau_{ij}$ as the time shift between the peaks of their accelerations. In practice, due to experimental noise, using just one time point (the peak) gives an unstable estimate. To calculate $\tau_{ij}$ in a robust way we must use the entire trajectories. This can be done by asking what is the delay $\tau_{ij}$ by which we have to time-shift the radial accelerations $a_j(t)$ to maximally overlap it with $a_i(t)$. This optimal shift corresponds to the time where the correlation function $G_{ij}(\tau)= \int dt\; {\bf a}_i(t)\cdot {\bf a}_j(t-\tau)$ reaches its maximum (inset).
{\bf e,} 
In the absence of experimental noise, for each triplet of birds, $i, j, k$ we must have, $\tau_{ik}+\tau_{kj} = \tau_{ij}$: if $i$ turns $20$ms before $k$, and $k$ turns $15$ms before $j$, then $i$ turns $35$ms before $j$ (Time Ordering Relation - TOR). Due to noise TOR will not hold strictly, but we still want it to be correct on average for $\tau_{ij}$ to make biological sense. We consider all triplets of birds and plot $\tau_{ik}+\tau_{kj}$ vs. $\tau_{ij}$. The data fall on the identity line with relatively small spread, confirming the temporal consistency of the turning delays.}
\label{fig:methods}
\end{figure*}

Here, we perform an experimental study of collective turns in natural flocks of starlings ({\it Sturnus vulgaris}).
Studying animal groups performing global turns is na\-tu\-ral for several reasons. First, turns are a paradigmatic example of collective change of state, often taking place under severe environmental constraints. Second, turns are  relevant from an adaptive perspective, as the mechanisms regulating global change of direction have an important anti-predatory value in many social species.  Third, collective turns can be defined sharply from a behavioural point of view and they are relatively easy to study empirically in natural conditions. 

We find that collective turns in starling flocks start from a few individuals and then propagate to the rest of the group like undamped sound waves, a phenomenon that the equations commonly adopted to describe collective motion fail to explain. We introduce a new theory mathematically identical to that describing superfluidity in liquid helium. The superfluid theory not only explains the observed sound-like propagation, but also predicts that the speed of propagation must be larger in more ordered flocks, a prediction confirmed by the experimental data. This result indicates that the efficiency of information transport during collective decision-making is quantitatively linked to the degree of behavioural polarization in the group.

\begin{figure*}[t!] 
  \centering
  \includegraphics[width=1.6 \columnwidth]{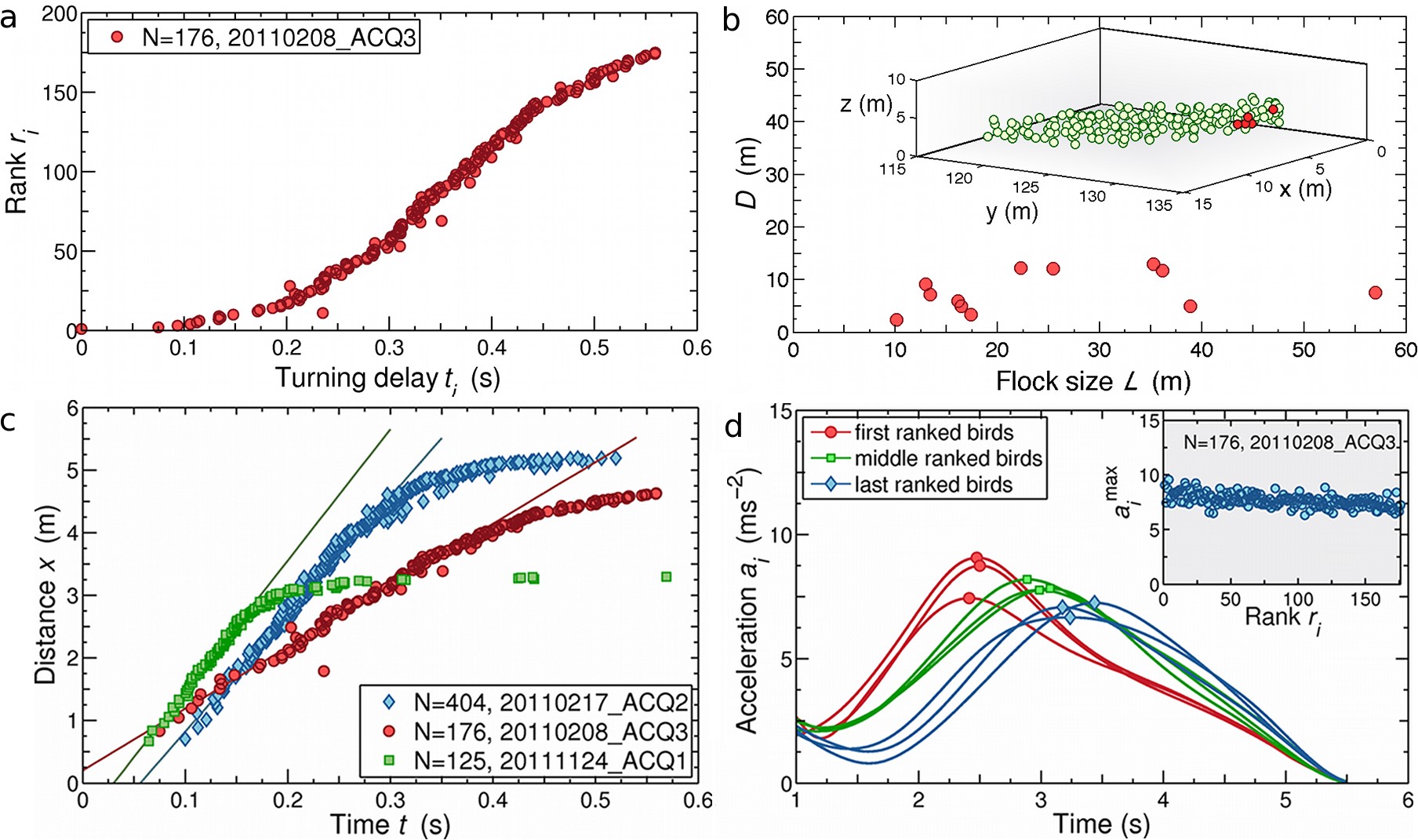}
 \caption{
{\bf Propagation of the turn across the flock.}
{\bf a,} 
The rank $r$ of each bird in the flock, i.e. its order in the turning sequence, is plotted vs its absolute turning delay $t$, i.e. the delay with
respect to the top bird in the rank (the first to turn). The convex toe of the curve for early times indicates that few birds take initially the decision to turn.
{\bf b,} 
The average mutual distance $D$ between the top $5$ birds in the rank does not increase with the linear size of the flock, $L$, hence indicating that the first birds to turn are actually close to each other in space. The result does not change if we use a different number of top birds. Inset: the actual position of the top $5$ birds (red) within a real flock.
{\bf c,} 
The distance $x$ traveled by the information in a time $t$ is proportional to the radius of the sphere containing the 
first $r(t)$ birds in the rank, namely $x(t)=[r(t)/\rho]^{1/3}$.
The linear regime of $x(t)$ allows us to define a `sound' speed of propagation, $c_s$, as the slope of $x(t)$ for early-intermediate times. The speed $c_s$ varies significantly from flock to flock (see also Table \ref{table:flocks}).
{\bf d,} 
The intensity of the peak of the radial acceleration, $a^\mathrm{max}$, (solid symbols) decreases very weakly in passing from 
the first to the last turning birds. In the inset, we plot $a_i^\mathrm{max}$ vs the rank $r_i$ for each bird.
This slow decay indicates that the information propagates through the flock with negligible attenuation.
 }
\label{fig:results}
\end{figure*}

We study natural flocks of starlings performing aerial display at dusk \cite{ballerini+al_08a,ballerini+al_08b}. By using a 3-cameras setup we reconstruct the full $3d$ dynamical trajectory of each bird in the flock. We have negligible time fragmentation: 90\% of the reconstructed trajectories last more than 90\% of the duration of the studied event. Cameras shoot at $170$Hz (Methods). From our pool of data we select 12 flocks, each one performing a collective turn.  Turns typically last a few seconds. In Fig.~\ref{fig:methods}
we present samples of the reconstructed trajectories.
When bird $i$ makes a turn, the modulus of its radial acceleration, $a_i(t)$, has a maximum (Fig.~\ref{fig:methods}-d).
We exploit this simple kinematic fact to organize all birds in the flock according to their temporal relationships.
For each pair of birds, $i$ and $j$, we calculate  their mutual turning delay, $\tau_{ij}$, namely the amount of time by which bird $j$ turns before ($\tau_{ij}>0$) or after ($\tau_{ij}<0$) bird $i$ (Fig.~\ref{fig:methods}-d,e). By using the delays $\tau_{ij}$ we then rank all birds in the flock according to their turning order, that is we find who is the first to turn,  who is second, and so on. In this ranking, each bird $i$ is labelled by its rank, $r_i$, and by its absolute turning time, $t_i$, which is defined as the delay with respect to the top bird in the rank, i.e. the first to turn (see Methods and Appendix \ref{delays} for details). To represent the ranking in a compelling way, we plot the rank $r_i$ of each bird as a function of its absolute turning time $t_i$, thus obtaining the ranking curve, $r(t)$ reported in Fig.~\ref{fig:results}-a.

By using the ranking we can find out whether the turn has extended or localized spatial origin. In the extended case, a large number of birds start to turn all at the same time, hence we would expect to find many birds packed in a very short time lag at the top of the ranking, i.e. a negative second derivative of $r(t)$ for small $t$. What we see from the data is the opposite: the ranking curve $r(t)$ is convex for early times, meaning that the turn is started by very few birds (Fig.~\ref{fig:results}-a). 
Moreover, we find that the first birds to turn (say, the top $5$ in the rank) are physically close to each other. More precisely, their average mutual distance $D$ does not scale with the size $L$ of the flock (Fig.~\ref{fig:results}-b). This result indicates that the number of birds initiating the turn is not proportional to the volume ($D\sim L^3$), nor to the surface ($D\sim L^2$) of the flock. Hence, the decision to turn has a spatially localized origin and it then travels across the flock through a transfer of information from bird to bird.

To understand how effective is this transport of information we need to calculate how much distance $x$ the information travels in a time $t$, i.e. we need the dispersion law. We are in three dimensions and the turn has a localized origin, hence $x(t)$ is equal to the radius of the sphere containing the first $r(t)$ birds in the rank, 
namely $x(t)=[r(t)/\rho]^{1/3}$, where $\rho$ is the density of the flock (Fig.~\ref{fig:results}-c). 
The most striking feature of the propagation curve $x(t)$ is that there is a clear linear regime for early-intermediate times (before border effects kick in). The distance traveled by the information grows linearly with time, $x(t) = c_s t $, just as a sound wave. The parameter $c_s$ is the speed of propagation of directional information, which is in the range $10-20$ meters per second (Table I).
The second important result is that the information to turn propagates across the flock with negligible attenuation (Fig.~\ref{fig:results}-d). This is nontrivial: flocks are large, the information to turn dynamically reaches all birds through a lot of intermediate passages, so that a substantial level of damping could be expected. Yet it is not so. This phenomenon too is reminiscent of sound propagation.

The speed of propagation of the information, $c_s$, varies significantly from flock to flock (Fig.~\ref{fig:results}-c and Table \ref{table:flocks}). It therefore seems that some flocks are more efficient than others to transport information. Why is that? Thinking about sound, one may naively expect the speed to depend on density. However, the variability of $c_s$ does not disappear by rescaling it with the flock's density. In fact, even though a linear dispersion law is suggestive of sound propagation, we should not forget that what propagates during the turn are fluctuations of {\it orientation}, not of {\it density}.  We shall see later that the variability of $c_s$ has an entirely different explanation.


Linear propagation and low damping of the signal are key factors in achieving an efficient collective decision of the group. Both sub-linear propagation and attenuation would result into a physical spread of the flock, and eventually into total disruption. Do current theories of collective motion account for such an efficient transport of directional information? 
Virtually all theoretical descriptions are based on alignment dynamics: each individual tends to keep its direction of
motion as close as possible to that of its neighbours \cite{huth_92, vicsek+al_95,toner+tu_98,couzin+al_02,gregoire+chate_04}, 
\begin{equation}
{\bf v}_i(t+1) = {\bf v}_i(t) + J\sum_{j\in i} {\bf v}_j(t)  \ ,
\label{rut}
\end{equation}
where the vector ${\bf v}_i$ is the velocity of bird $i$ and the sum extends over all neighbours $j$ of $i$ (be they metric or topological \cite{ballerini+al_08a,ginelli+chate_10}). We have disregarded noise/temperature, which is inessential for what follows; we will just assume that the alignment strength $J$ is large, so that we are in the deeply ordered phase (as natural flocks are \cite{cavagna+al_10}).  In continuous time equation (\ref{rut}) is equivalent to a zero-temperature Langevin equation, 
\begin{equation}
\frac{d {\bf v}_i}{dt} = - \frac{\partial H}{\partial{\bf  v}_i}  \quad \quad , \quad \quad H= - J \sum_{\langle ij\rangle} {\bf v}_i \cdot {\bf v}_j \ .
\label{ham1}
\end{equation}
According to these relations each bird updates its velocity following a {\it social force}, ${\bf F}^i_\mathrm{s}=- {\partial H}/{\partial {\bf v}_i}$, produced by its neighbours. The Hamiltonian $H$ in  (\ref{ham1}) is the same as that of a ferromagnetic system, where the birds velocities ${\bf v}_i$ play the role of magnetic spins \cite{bialek+al_12}.

To simplify the algebra we exploit the fact that the trajectories of birds during a turn lie approximately on a plane (Fig.~\ref{fig:methods}-b,c). This allows us to use a two-dimensional velocity, ${\mathbf v}_i=(v_i^x,v_i^y) = v \, e^{i\varphi_i}$, where the phase $\varphi_i$  is the angle between the direction of motion of $i$ and that of the flock (we make the standard assumption that $v$ is constant). In the highly ordered phase the velocities ${\bf v}_i$ differ little from the collective one, so that $\varphi_i\ll1$. We can thus expand $H$ in eq.(\ref{ham1}) \cite{bialek+al_12},
\begin{equation}
H= \frac{J}{2}\sum_{\langle ij\rangle} (\varphi_i - \varphi_j)^2  
= 
\frac{1}{2}a^2J \int \frac{d^3x}{a^3} \, \left[ {\boldsymbol\nabla} \varphi(x,t)\right]^2
\ ,
\label{align}
\end{equation}
where $a$ is the average nearest neighbours  distance and a term $v^2$ has been reabsorbed into $J$.
The Langevin equation associated to Hamiltonian (\ref{align})  is,
\begin{equation}
\frac{\partial \varphi}{\partial t} = -\frac{\delta H}{\delta \varphi}  =a^2 J\, \nabla^2\varphi
\ .
\label{diffusion}
\end{equation}
Relation (\ref{diffusion}) is a diffusion equation for the phase $\varphi$, and it has dispersion law $\omega = i k^2$.
This result has two consequences, both in sharp contrast with the empirical data: i) information travels much slower than linearly,  $x\sim \sqrt{t}$, at variance with the linear propagation we find in turning flocks; ii) the frequency is imaginary, meaning that this is a non-propagating mode. Transfer of information gets damped exponentially in space and time, again in stark disagreement with the brisk, undamped propagation we observe in flocks. 


The standard theory has two problems. 
First, it seems to be missing some conservation law. Hamiltonian (\ref{ham1}) is invariant under a global gauge symmetry, namely the uniform rotation of the velocities ${\bf v}_i$ ($\varphi_i\to\varphi_i +\delta\varphi$). This symmetry encodes the fact that all directions of flight are equivalent for a flock. Through Noether's theorem, a symmetry implies in general a conservation law, of which, however, there is no trace in the standard theory. A hidden conservation may heavily affect the dispersion law, because a conserved quantity cannot be relaxed locally, but it must be transported across the system. Second, equation (\ref{diffusion}) completely neglects behavioural inertia, as the social force, $F_\mathrm{s}= a J\nabla^2\varphi$, controls directly $\dot \varphi$, rather than $\ddot\varphi$. This is odd: imagine that the interaction with the neighbours requires bird $i$ to perform a U-turn in one time step. This behaviour is allowed by the standard theory, although it is clearly unreasonable.

\begin{figure*}[t!] 
  \centering
  \includegraphics[width=1.7\columnwidth,]{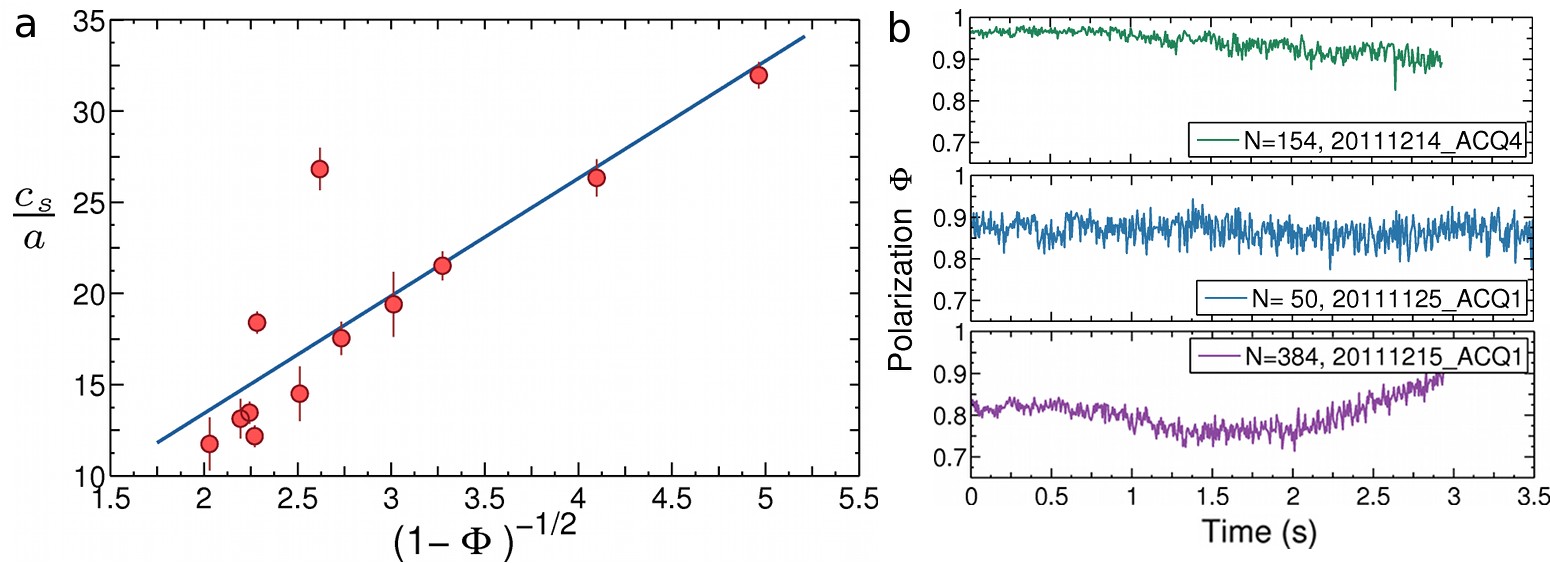}
 \caption{
{\bf Superfluid prediction.}
{\bf a}, 
The superfluid theory predicts that the rescaled speed of propagation of the turn, $c_s/a$, must be a linear function of $1/\sqrt{1-\Phi}$, where $\Phi$ is the polarization. The prediction is verified by the empirical data (P-value: $P=3.1\times10^{-4}$; correlation coefficient: $R^2=0.74$). Each point is a different turning flock. Error bars on $c_s$ are obtained from its variability under changing the linear fitting regime of $x(t)$. 
$c_s/a$ has the dimensions of sec$^{-1}$. 
The slope of this line is equal to $\sqrt{\epsilon/\chi}$ - equation\eqref{secondsound}. 
{\bf b}, 
Polarization as a function of time in three different flocks. The value of $\Phi$ reported in panel {\bf a} corresponds to the time average over the entire duration of the turn.
}
\label{fig:prediction}
\end{figure*}

To address these problems we follow Landau's approach \cite{LG}, namely we identify a suitable order parameter and write the simplest Hamiltonian compatible with the symmetries and the constraints of the system. The phase $\varphi$ is the obvious order parameter. The gauge symmetry implies that $H$ can only be a function of ${\boldsymbol\nabla}\varphi$, not of $\varphi$ itself, as expressed by \eqref{align}. On the other hand, the inertial constraint requires a kinetic term, $s_z^2/2\chi$, where $s_z$ is the canonical momentum conjugated to $\varphi$, and $\chi$ is the generalized moment of inertia. We therefore propose the novel Hamiltonian,
\begin{equation}
H= \int \frac{d^3x}{a^3} \left\{
 \frac{1}{2} \rho_s   \left[ {\boldsymbol  \nabla} \varphi(x,t) \right]^2
 +
 \frac{s_z^2(x,t)}{2\chi}
 \right\} \ ,
 \label{nobu}
\end{equation}
where $\rho_s\equiv a^2 J$, is the rescaled alignment coupling constant, or stiffness \cite{fisher_73}.
The momentum $s_z$ is defined as the local generator of the rotations parametrized by the phase $\varphi$, so that $(s_z,\varphi)$ are generalized action-angle
canonical variables. It can be shown that $s_z$ is essentially the inverse radius of curvature of the trajectory, whereas $\chi$ is the behavioural resistance of a bird to change its instantaneous radius of curvature when a social force is exerted by its neighbours (see Appendix~\ref{spin} and \ref{inertia}).
The canonical equations of motion generated by (\ref{nobu}) are,
\begin{equation}
\frac{\partial \varphi}{\partial t} = \frac{\delta H}{\delta s_z} = \frac{s_z}{\chi} 
\quad ; \quad 
\frac{\partial s_z}{\partial t} = -\frac{\delta H}{\delta \varphi} =  \rho_s\, \nabla^2\varphi \ .
\label{canonical}
\end{equation}
The crucial consequence of the gauge symmetry is that the r.h.s of the second equation of motion is in fact a gradient, $ \nabla^2\varphi={\boldsymbol  \nabla \cdot \boldsymbol  \nabla }\varphi$, so that we can rewrite this relation as a continuity equation for $s_z(x,t)$, 
\begin{equation}
\frac{\partial s_z}{\partial t} - {\boldsymbol  \nabla} \cdot {\bf j}_z = 0 \ ,
\label{continuity}
\end{equation}
with current ${\bf  j}_z(x,t)=\rho_s \,{\boldsymbol \nabla} \varphi(x,t)$. We therefore find a conservation law.
Imagine that a strong misalignment among a subgroup of birds forms in a certain position of the flock. This causes a local excess of curvature, and thus an excitation of the field $s_z(x,t)$. Conservation law (\ref{continuity}) states that such excitation cannot be locally {\it dissipated} out, but it must be {\it transported} away. This mechanism, which is the essence of the new theory, gives rise to an undamped sound-like mode. Indeed, by taking the second derivative with respect to time in (\ref{canonical}) we get, 
\begin{equation}
\frac{\partial ^2\varphi}{\partial t^2} = c_s^2 \; \nabla^2\varphi  \quad \ , \quad c_s^2 = \rho_s/\chi\ .
\label{dalembert}
\end{equation}
Relation \eqref{dalembert} is D'Alembert's equation, describing waves propagating with speed $c_s$ and no damping. Its dispersion relation is linear, $\omega=c_s \, k$, which in terms of the distance $x$ traveled by the information in a time $t$, reads, $x=c_s\, t$. This is precisely the linear and undamped propagation law that we find in turning flocks.

Some flocking theories have established a correspondence between flocks and magnets, with birds velocities ${\bf v}_i$ playing the role of spins trying to align to each other \cite{vicsek+al_95, toner+tu_98, bialek+al_12}. However, up to now spins were virtual, as they did not obey proper Poisson rules. Within the present description, things change.
The fact that $s_z$ generates the symmetry parametrized by the phase $\varphi$ is expressed by Poisson relation, $\{{\bf v},s_z\}=\partial {\bf v}/\partial \varphi=i{\bf v}$, which in components reads,
\begin{equation}
\{v_x, s_z\} = \frac{\partial v_x}{\partial \varphi} = - v_y \ \ ; \ \
\{v_y, s_z\} = \frac{\partial v_y}{\partial \varphi} = v_x \ .
\label{poisson}
\end{equation}
If we interpret $v_x$ and $v_y$ as the $x,y$ components of the spin, equations (\ref{poisson}) show that $s_z$ is a {\it true} spin, namely the generator of the rotation in the space of the order parameter ${\bf v}$. This is the most general and fundamental definition of spin \cite{pauling+wilson_35}. Accordingly, ${\bf  j}_z = \rho_s {\boldsymbol \nabla} \varphi$ is the spin current and $\rho_s$ the spin stiffness \cite{halperin_69, halperin_76}. 

The theory of collective motion that we have introduced above is exactly the same as that describing superfluid liquid helium (He-II). It has been demonstrated long ago by Matsubara and Matsuda \cite{matsubara+matsuda_56, matsubara+matsuda_56b} that the lattice-gas model for Bose condensation in He-II is mathematically equivalent to the planar ferromagnetic model defined by equation (\ref{nobu}). This is not just a bizarre coincidence. Superfluidity is nestled into the identical mathematical structure of these apparently very different systems. The keystones of superfluidity are: i) existence of a gauge symmetry (arbitrariness of the quantum phase/arbitrariness of the flock direction of motion); ii) spontaneous symmetry breaking, i.e. emergence of a nonzero order parameter, ${\boldsymbol \psi}=|\psi|\, e^{i\varphi}$ (nonzero Bose wave function/nonzero flock's velocity); iii) coupling of the phase $\varphi$ to the generator of the gauge symmetry $s_z$ (Bose particle density/spin density). Irrespective of the physical and biological details, these three elements alone generate the dissipationless propagating mode described by equation (\ref{dalembert}), that is superfluidity  \cite{halperin_69, halperin_76, halperin_77, sonin_10}. In liquid helium, ${\bf j}_z$ is the current of the superfluid component and the propagating mode is called `second-sound'  \cite{helium_47, halperin_69}. In flocks, ${\bf j}_z$ transports spin, that is curvature, giving rise to the collective turn. 

The superfluid theory not only provides an explanation for the linear and undamped propagation of information in natural flocks, but it also makes a prediction about the dependence of the speed of propagation on the experimentally accessible quantities, thus making sense of the otherwise unexplained variability of $c_s$ from flock to flock. We recall that $c_s^2=\rho_s/\chi$, where the stiffness is $\rho_s=a^2J$.
The alignment strength can be written as $J= \epsilon/(1-\Phi)$, where $\Phi$ is the polarization, namely the degree of alignment in the flock, $\Phi = ||(1/N) \sum_i {\bf v}_i/v_i ||$, and $\epsilon$ is an energy constant, setting the scale of the alignment interaction \cite{bialek+al_12}. We therefore obtain,
\begin{equation}
c_s = \sqrt{\epsilon/\chi} \; \frac{a}{\sqrt{1-\Phi}}  \ .
\label{secondsound}
\end{equation}
Equation (\ref{secondsound}) predicts that the speed of propagation of the turn across a given flock must be larger the larger the degree of alignment $\Phi$  in that flock. We report $c_s/a$ vs. $1/\sqrt{1-\Phi}$ for all our flocks in Fig.~\ref{fig:prediction}. Data show a clear linear dependence, exactly as predicted by equation (\ref{secondsound}). 
We remark that the square root behaviour reproduced by the data is nontrivial: the polarization is a dimensionless quantity, hence the functional dependence of $c_s$ on $\Phi$ cannot be worked out by mere dimensional analysis. Interestingly, equation \eqref{secondsound} also implies that the only empirically observable quantity is the ratio between the scale of the alignment interaction, $\epsilon$, and the turning inertia $\chi$, so that different species (or artificial entities) may have very different values of $\epsilon$ and $\chi$, but still have a comparable information transport efficiency.

Superfluid propagation of information in collective motion has never been
discovered before, not even in models more realistic than \eqref{rut},
as those studied in  \cite{krishna_04, szabo_09, gautrais+al_09, hemelrijk_11}. In fact, 
as we have seen, superfluidity stems from associating inertial terms to the fundamental conservation law 
generated by the gauge symmetry. It would be interesting to reconsider flocking models 
with inertia  in the light of the present results.
Moreover, we note that, although the dynamical equation (\ref{dalembert}) is different from (\ref{diffusion}), the {\it static} properties of the field $\varphi(x,t)$, and in particular its equal time correlations \cite{bialek+al_12}, are the same as those described by Hamiltonian (\ref{align}). This result is due to the separation between coordinates and momenta.


The link between speed of propagation of the information, $c_s$, and behavioural polarization, $\Phi$,  is not an evolutionary trait, but the mathematical consequence of the gauge symmetry. However, the specific level of polarization of a flock is not fixed by math, nor by symmetry, but by adaptive factors. In many social species polarization is very large \cite{krause+ruxton_02, camazine+al_01, couzin+krause_03, cavagna+al_10}. Global order is indeed the most conspicuous trait of collective behaviour. However, were the only concern of a bird not to bump into its neighbours, such a large polarization would be difficult to justify. Flocks for which we have data are rather diluted systems, with packing fraction lower than $0.01$ \cite{cavagna+al_08b}. Yet these same flocks are very stiff, precisely in the superfluid sense: alignment strength, $\rho_s\sim J$, is large and polarization is close to $1$. Why is that?
In collective decision-making swift transport of information is beneficial to the group. In the case of turns this is obvious: during the turn the wavefront divides approximately the flock into two groups of birds with different directions of motion. Such misalignment causes a spatial spread of the flock, with overall loss of cohesion. The slower the speed $c_s$ of the wavefront, the more severe this loss. It is reasonable to believe that this is a general mechanism. Every collective decision drives the group through a momentary lapse of cohesion, due to the transient coexistence of different behavioural states. The superfluid link between high behavioural polarization and fast propagation of the information suggests that keeping this lapse to a minimum, therefore achieving a swift and robust collective decision, may be the adaptive drive for the high degree of order observed in living groups.


We thank Emmanuele Cappelluti, Claudio Castellani, Jos\`e Garcia Lorenzana for helpful discussions, and Pasquale Calabrese for many useful comments on the manuscript. We also acknowledge the advice of Carlo Lucibello on tracking and of Edmondo Silvestri on segmentation.
This work was supported by grants IIT--Seed Artswarm, ERC--StG n.257126 and AFOSR--Z809101.



\appendix

%
%


\section{The spin}
\label{spin}
Let us define {\it external} space the space of the birds coordinates and {\it internal} space (or target space) the space of the order parameter, 
namely the planar velocity. Indeed, ${\bf v}(x,t)$ is a map between the external space $\mathbb{R}^3\times\mathbb{R}$ and the internal space SO$(2)$ (the circle). It is essential to understand that the phase $\varphi$ parametrizes rotations in the internal space of the velocity and it must not be confused with the standard angle $\theta$ of $2d$ polar coordinates, which parametrizes rotations in the external space of positions. The easiest way to understand this is the following: a bird flying straight (with respect to an arbitrary fixed reference frame - that of our cameras, for example) has $\dot \varphi =0$, but $\dot\theta\neq 0$.

It is interesting to note that rotations parametrized by these two angles, $\varphi$ and $\theta$, correspond to two very different types of collective turns. A rotation parametrized by $\varphi$ corresponds to an {\it equal radius} turn, i.e. a turn in which all birds have (approximately) the same radius of curvature and where trajectories cross. This is exactly the way adopted by flocks to turn (see Fig.~\ref{fig:methods}),
and it has a clear biological motivation: it keeps the speed $v$ (approximately) constant throughout the flock. Equal radius turning was first experimentally discovered in \cite{heppner_92} (see also \cite{ballerini+al_08b}). On the other hand, a rotation parametrized by $\theta$ corresponds to a {\it parallel path} turn, typical of rigid bodies. This kind of turn implies different radii of curvature for different points, and therefore it requires an increase of speed of birds on the outer side of the turn, which is clearly biologically unreasonable.

The generator of the external $\theta$-rotations is the standard angular momentum, $l_z$, whereas the generator of the internal $\varphi$-rotations is the spin, $s_z$, which is conserved by the continuity equation. In order to have some intuition about the physical nature of $s_z$ we must connect external to internal spaces. This connection is established by the kinematic equation,
\begin{equation}
\dot {\bf x} = v\, e^{i\varphi}  \ ,
\label{external}
\end{equation}
expressing the simple fact that birds are not anchored to a lattice, but they follow their velocity vectors. If we consider the speed $v$ approximately constant, equation (\ref{external}) implies,
\begin{equation}
\dot\varphi = v/R  \ ,
\label{barrio}
\end{equation}
where $R$ is the instantaneous radius of curvature. Using (\ref{barrio}) together with (\ref{canonical}), we get,
\begin{equation}
\ddot\varphi = - \frac{v}{R^2} \dot R = \frac{\dot s_z}{\chi} \ ,
\end{equation}
and dividing by the first equation in (\ref{canonical}) we finally obtain,
\begin{equation}
\frac{\dot s_z}{s_z} = - \frac{\dot R}{R}  \  .
\label{rebibbia}
\end{equation}
This equation connects the rather abstract internal spin, $s_z$, to a very clear kinematic quantity, the radius of curvature $R$. By integrating (\ref{rebibbia}), we have,
\begin{equation}
s_z \sim 1/R  = \kappa\ ,
\end{equation}
where $\kappa = 1/R$ is, by definition, the {\it curvature} of the trajectory. 
Therefore, once the connection with the external space is performed, the spin turns out to be essentially the curvature. This is why a bird flying straight ($R=\infty, \kappa\sim 0$) has $s_z\sim0$, while it has nonzero standard angular momentum $l_z$. A change (in time) of the spin $s_z^i$ of bird $i$, due to the social force exerted by the neighbours of $i$, corresponds to a change (in time) of its instantaneous radius of curvature, $R_i$, and curvature, $\kappa_i$. Hence, what actually propagates across the flock during the turn is a fluctuation (in space and time) of the curvature field, $\kappa(x,t)$. Before the turn, the flock is flying almost straight, $R \gg 1, \kappa\sim 0, s_z\sim 0$. Then the turns sparks in some part of the flock, causing an increase of the curvature $\kappa$, i.e. an increase of $s_z$. This change sweeps through space and time until the whole flock has turned. Finally, after the turn, the flock relaxes back to $R\gg1, \kappa\sim 0, s_z\sim0$.
Mathematically, this propagation of the curvature, i.e. of $s_z$, derives from the canonical equations \eqref{canonical}: by taking the second derivative with respect to time one obtains a D'Alembert wave equation for $s_z(x,t)$ identical to that obeyed by $\varphi(x,t)$, eq.\eqref{dalembert}.

Conservation law (\ref{continuity}) states that the spin-curvature, $s_z(x,t)\sim \kappa(x,t)$, obeys a continuity equation. As we have seen, this conservation law is crucial to determine sound-like propagation. Continuity means that the trajectory curvature in a given volume cell of the system cannot change unless it is transported into, or out of, that cell by a spin current, ${\bf  j}_z=\rho_s{\boldsymbol\nabla} \varphi$. We can reformulate this by saying that, if a certain excess of curvature, namely a strong misalignment among a certain group of individuals, forms in a given point of the system, it cannot be simply dissipated out. Rather, such excitation creates a social force that makes the neighbours turn, and their neighbours too, and so on, so that the excess curvature is transported away, instead of dissipated.

Finally, note that the spin $s_z$ is {\it not} the $z$ component of the velocity. Also note that the rotation generated by $s_z$ is the very transformation under which Hamiltonian (\ref{nobu}) is symmetric. 


\section{The generalized moment of inertia}
\label{inertia}

As we have seen, the spin $s_z$ is not the standard angular momentum $l_z$. Accordingly, the generalized moment of inertia $\chi$, is not the standard moment of inertia, which
in the case of circular motion is, $I=m R^2$, where $m$ is the mass. So, what is the physical and biological meaning of $\chi$? From the canonical point, the answer is clear: $\chi$ is the inertia to changing $\dot\varphi$. Indeed, equation (\ref{dalembert}) can be rewritten as,
\begin{equation}
\chi = \frac{a F_s}{\ddot \varphi} \ ,
\label{kong}
\end{equation}
where $F_s=aJ\nabla^2\varphi$ is the social force exerted by the neighbours. Hence, the generalized moment of inertia $\chi$ is {\it defined} as the ratio between
the social force (the cause) and the change of angular velocity (the effect). This is the standard definition of inertia: the ratio between force (cause) and acceleration (effect). 
However, in this case $F_s$ is a generalized (social) force, and $a F_s$ is a generalized torque, hence $\chi$ is not the standard moment of inertia.

To better grasp the biological meaning of $\chi$ we must, once again, bridge the gap between internal and external space. By using equations (\ref{external}), (\ref{barrio}) and (\ref{kong}) we obtain several cause-effect relations shading light on the physical and biological meaning of $\chi$. The first relation connects the social force to the change of radius $R$, or equivalently to the change of curvature $\kappa$, 
\begin{equation}
\chi = - \left( \frac{R^2}{v} \right) \; \frac{a F_s}{\dot R}  =  \left( \frac{1}{v} \right) \; \frac{a F_s}{\dot \kappa}
\ .
\label{pongo}
\end{equation}
Hence, $\chi$ is the resistance of a bird to change its instantaneous radius of curvature $R$  (the effect), when a given social force $F_s$ (the cause) is exerted. 

Another interesting relation can be obtained in terms of the banking angle $\gamma$. A banked turn \cite{norberg_90,videler_05} is the typical way birds (and planes) change their heading. It consists in a gentle roll, so to form an angle $\gamma$ between the axis of the wings and
the horizontal plane. In this way, part of the total lift goes into a centripetal force, $F_c = m g \gamma$, making the bird turn ($m$ is the mass, $g$ the gravitational acceleration and $\gamma \ll 1$). From (\ref{pongo}), it is straightforward to prove that,
\begin{equation}
\chi = \left( \frac{v}{g} \right) \; \frac{a F_s}{\dot \gamma}  \ .
\label{ponio}
\end{equation}
According to (\ref{ponio}), the generalized moment of inertia $\chi$ is the resistance of a bird to change its banking angle $\gamma$. Here, $\dot\gamma$ is the effect of the social force, $F_s$, and $\chi$ sets the ratio between cause and effect. Notice that equations (\ref{pongo}) and (\ref{ponio}) are clearly non-canonical definitions of the inertia $\chi$, because at the denominator they both have a first order derivative in time, rather than a second order one, as in the canonical equation (\ref{kong}).

Let us emphasize once again that $\chi$ is {\it not} the standard, mechanical moment of inertia, $I=mR^2$, but rather a social, or sensorimotor, resistance of the bird to change $R$ or $\gamma$. It is not possible to write an {\it ab inicio} expression for $\chi$ in terms of primary mechanical quantities, like mass, radius, etc. To understand this fact, let us imagine that at some 
point the neighbours of bird $i$ all sharpen their banking angle $\gamma$, thus creating a strong social force, $F_s=aJ\nabla^2\varphi$, acting on $i$. What we call a social `force' is
in fact a shortcut to describe a very complex sensorimotor process: a nonzero $\nabla^2\varphi$ means that some of the neighbours of $i$ are now about to crash into $i$. This is most likely perceived by $i$, which decides to change its own $\gamma$ and make it equal to that of the neighbours, thus avoiding the crash. However, the degree by which $i$ will react to the imminent crash, or conversely the resistance to this reaction (which is $\chi$), is the result of a {\it very} complex trade-off. Let us analyze this trade-off by pretending to be $i$.

On one side, there is the price of the crash. How imminent is it? This will depend on both the nearest neighbour distance and on the mutual velocity. How bad would that be? Perhaps, I can ignore my neighbours, and just change them, without any real crash. How confident I am into my capability to change $\gamma$? If I am very agile, I can wait a bit longer before changing my $\gamma$. On the other side, there is the price to changing $\gamma$. How much will it cost me energetically? By increasing $\gamma$ I will increase the drag against air, otherwise I fall down. But to do this I must increase the power, which is very costly. Can I manage?

The generalized moment of inertia $\chi$ is the very final product of this process. Clearly, we cannot know a priori its value. But we can define it, and measure it. This is exactly what we have done by measuring the speed of propagation of the turn across the flock. In this sense, the situation is the same as in real magnetic systems \cite{halperin_69}: the parameter $\chi$ is the magnetic susceptibility to an external field, which cannot be simply expressed as a function of the microscopic parameters of the theory, but it can be experimentally measured.


\section{Dimensional analysis}
\label{dim-an}

In the superfluid theory we have an Hamiltonian that is the sum of two parts. Hence, we have to be careful
with physical dimensions.
The phase  is of course a pure number, $[\varphi]=[1]$, whereas the alignment coupling constant has the dimensions of an energy, $[J] = [e]$. 
In this way the social force has the dimensions of a true force, $[F_s] = [aJ\nabla^2 \varphi] = [e\cdot x^{-1}]$ and the spin has the dimensions of an angular 
momentum, $[s_z]=[e\cdot t]$, i.e. of an action. Accordingly, $\chi$ has the dimensions of a true moment of inertia, $[\chi]=[e\cdot t^2]$.
Notice that the term appearing in equation (\ref{canonical}) is $a^2 J \nabla^2\varphi= a \,F_s$, which is dimensionally a torque.
Hence, the derivative of an angular momentum is a torque, as it should. 

By definition, the polarization is a pure number, $[\Phi]=[1]$. This is why in the relation
linking alignment coupling constant to polarization,   $J= \epsilon/(1-\Phi)$, we need a dimensional constant with the dimensions of an energy, $[\epsilon]=[e]$. As we have said, $\epsilon$ sets the scale of the alignment interaction. Roughly speaking, $\epsilon$ (as well as $\chi$) is what distinguishes species $A$ from species $B$, or bird flocks from aircraft formations. Finally, with to the above physical dimensions, the speed of propagation of information across the flock, $c_s$, is measured in meters per second, as it should.


\section{General off-plane case}
\label{modelg}

Our initial assumption that the birds' velocities lie on a plane during the turn, namely that the turn has very small torsion, although experimentally satisfied (see Fig.1b,c), it is not at all a necessary condition for our mathematical description. The most general formulation of our result holds even with a truly $3d$ order parameter ${\bf  v}_i$ \cite{halperin_69}. 

If we assume that the mean direction of motion of the flock points in the $x$ direction, then there will be full $3d$ fluctuations of the individual velocities ${\bf v}_i$ around the overall direction of motion of the flock, generating small components of ${\bf v}_i$ along the two orthogonal axes, $z$ and $y$. Therefore, we must define two phases, $\varphi_z$ and $\varphi_y$ and the Hamiltonian can be spin-wave expanded in terms of these two fields. The phase $\varphi_z$ parametrizes rotations of ${\bf v}_i$ around the $z$ axis (as in the planar - zero torsion case), whereas $\varphi_y$ parametrizes rotations of ${\bf v}_i$ around the $y$ axis. In this fully $3d$ case the Hamiltonian is given by \cite{halperin_69},
\begin{equation}
H= \int \frac{d^3x}{a^3} 
 \frac{1}{2} \rho_s
 \left[
   \left( {\boldsymbol \nabla} \varphi_z \right)^2
   +
      \left( {\boldsymbol \nabla} \varphi_y \right)^2
\right] 
+ \frac{1}{2\chi}
 \left[
s_z^2 +s_y^2
 \right]  \, ,
\label{ham3d}
\end{equation}
where $\rho_s=a^2 J$ is, as usual, the stiffness. The equations of motion are,
\begin{eqnarray}
\frac{\partial \varphi_{\alpha}}{\partial t} &=& \frac{\partial H}{\partial s_{\alpha}} = \frac{s_{\alpha}}{\chi} \ ,
\label{canonical1-yz}
\\
\frac{\partial s_{\alpha}}{\partial t} &=& -\frac{\partial H}{\partial \varphi_{\alpha}} =  a^2 J \, \nabla^2\varphi_{\alpha} = {\boldsymbol \nabla} \cdot {\bf j}_{\alpha} \, ,
\label{canonical2-yz}
\end{eqnarray}
with ${\alpha}=y, z,$ giving rise to two D'Alembert equations,
\begin{equation}
\frac{\partial ^2\varphi_{\alpha}}{\partial t^2} - c_s^2 \; \nabla^2\varphi_{\alpha} = 0 \quad \ , \quad c_s^2 = \rho_s/\chi\ .
\label{dalembert-yz}
\end{equation}
In the full 3$d$ case we therefore obtain {\it two}, rather than one, propagating dissipationless modes, 
along the transverse directions $y$ and $z$. This is just a manifestation of Goldstone's theorem \cite{goldstone_61}.

These equations are exactly the same as for model G in the 
Halperin-Hohenberg classification of dynamical universality classes \cite{halperin_69,halperin_77}.
Model G does not describe superfluid liquid helium, but an isotropic Heisenberg antiferromagnet with 
staggered magnetization as a non-conserved order parameter, and total magnetization
as a constant of motion. 
An experimental realization of a $3d$ isotropic antiferromagnet is $\mathrm{RbMnF}_3$, a compound  
whose  dynamics is characterized by the transverse spin-wave modes \eqref{dalembert-yz}. Notice that
also in this system there is superfluid transport. As discussed in the main text, superfluidity is not restricted to
liquid helium II, but it is rather built into the mathematical details of the theory. In particular, it is the product of
symmetry and conservation laws. In the full 3$d$ case described here (model G) these ingredients give rise 
to superfluid transport exactly as in the planar (He-II) case.

To write \eqref{ham3d} and \eqref{dalembert-yz} 
we have assumed that the two excitations $\varphi_z$ and $\varphi_y$ are equally likely, so that the only symmetry breaking direction is that of motion. In fact, recent studies on individual diffusion in starling flocks show that gravity is another symmetry breaking direction, heavily suppressing  fluctuations along the vertical plane \cite{duarte_13}. If we identify $z$ with the axis of gravity, this suppression would imply that rotations of the velocity around the $y$ axis are suppressed, and therefore that $\varphi_y$ is less relevant a degree of freedom than $\varphi_z$. This  suppression induced by gravity is likely the cause of the planar-like turns we observe in flocks and it thus justifies the adoption of the simpler planar description of the main text.


\section{Dissipation}
\label{dissipation}

The diffusive equation of motion \eqref{diffusion}  derived from the standard theory can be obtained as the over-damped limit of the new theory, once we introduce a dissipative term proportional to $\dot\varphi$ in the  equation of motion,
\begin{equation}
\chi \frac{\partial ^2\varphi}{\partial t^2} + \eta  \frac{\partial \varphi}{\partial t} - \rho_s \; \nabla^2\varphi = 0 \ ,
\label{dalembert2}
\end{equation}
with $\rho_s = a^2 J$ and where $\eta$ is a generalized friction coefficient. From this we get,
\begin{equation}
\chi \omega^2 - i \eta \omega -\rho_s k^2 =0\ .
\end{equation}
In the limit $\eta \gg \chi$ we simply get the diffusive result, $\omega = i (\rho_s/\eta) k^2$. In general, however, we obtain,
\begin{equation}
\omega = i\frac{\eta}{2\chi}\pm c_s k \sqrt{1- k_0^2/k^2} \ ,
\end{equation}
where, as usual, the propagation speed is $c_s=\sqrt{\rho_s/\chi}$ and,
\begin{equation}
k_0 \equiv \frac{\eta}{2\sqrt{\rho_s\chi}} \ .
\end{equation}
If we define the dissipation time scale, $\tau \equiv 2\chi/\eta$, and the zero-dissipation frequency, $\omega_0\equiv c_s k$, we can rewrite the dispersion law as,
\begin{equation}
\omega = i/\tau \pm \omega_0 \sqrt{1- k_0^2/k^2} \ .
\end{equation}
With zero dissipation, we get $k_0=0$, $\tau=\infty$ and $\omega = \omega_0$. For $\eta\neq 0$, on the other hand,
we have two regimes, according to the value of the friction coefficient and of the wave number $k$. 
For $k\geq k_0$ we have {\it attenuated} propagating waves, as the frequency has both a real and an imaginary part.
For $k < k_0$ we have {\it evanescent} waves: the frequency is purely imaginary, there is no propagation, but pure exponential decay.

The smallest value of $k$ in the system is $k_\mathrm{min} \sim 1/L$, where $L$ is the linear size of the flock. 
Hence, small dissipation is defined by the relation,
\begin{equation}
\eta < \frac{\sqrt{\rho_s\chi}}{L} \quad :
\quad
\mathrm{small\  dissipation}
\ .
\label{smammal}
\end{equation}
With small dissipation there is linear propagation of the information and the time scale of the exponential decay is set by
$\tau = 2\chi/\eta$. From \eqref{smammal} we get,
\begin{equation}
\tau >  \sqrt{\chi/\rho_s} \ L = L/c_s \ .
\end{equation}
Therefore, small dissipation implies that the damping time constant is larger than the time the information employs to travel across the  flock. In other words, the signal is effectively very weakly damped across the length scale of interest. We conclude that even when a small dissipation is present, propagation of information is qualitatively the same as that described by the zero dissipation theory.


\section{Mutual delay vs reaction time}
\label{delays}

One may think that the mutual delay between two birds, $\tau_{ij}$, is  the same as the reaction time, $\tau_\mathrm{R}$, namely the time between the stimulus provided by $j$ and the consequential action of $i$. However, this is not the case.

Let us assume $j$ is the first bird to turn, and that $i$ is second. By definition of reaction time, $i$ begins its turn $\tau_\mathrm{R}$ seconds after $j$. However, we do not define $\tau_{ij}$ as the difference between the starting instants of the two turns: there is no practical and robust way to do that, because each birds turns smoothly, so that it is impossible to define the `start' of the turn. To compute $\tau_{ij}$ we use the {\it entire} trajectory of both birds, by finding the time shift that maximally overlaps the accelerations of $i$ and $j$ (see Fig.1). If the function $a_i(t)$ were {\it exactly} the same as the function $a_j(t-\tau_{ij})$, then we would have $\tau_{ij} = \tau_\mathrm{R}$. This, however, is never the case. First of all there is noise, so that the two curves, $a_i(t)$ and $a_j(t)$, are only approximately shifted with respect to each other. But more importantly, the second bird $j$ can try to `catch up' during the turn, so that the delay at the end of the turn is {\it shorter} than the delay at the beginning of the turn, which is the reaction time.
In this case, the delay $\tau_{ij}$ would be a value intermediate between those two times, hence giving a value smaller than $\tau_\mathrm{R}$. The opposite can happen too: bird $i$ could in fact lose ground during the turn, so that the delay at the end of the turn is {\it longer} than the reaction time, and $\tau_{ij}$ is larger than $\tau_\mathrm{R}$.

There is, however, a robust way to extract a timescale from our data, and to compare it to standard reaction times in birds. The speed of propagation, $c_s$, is expressed in meters per second, so that $c_s/a$ is the inverse of a time constant: it is essentially the time the information needs to travel the nearest neighbour distance, $a$. From Fig.3 we see that 
$a/c_s$ ranges between $25$ms up to $100$ms, with an average around $50$ms. Again, this is not the reaction time, as it depends on the information transfer mechanisms. However, it is reassuring to see that $50$ms is in the physiological range of reaction times for birds in general and for starlings in particular \cite{heppner_77}.


\begin{figure*}[t!]
  \centering
  \includegraphics[width=2\columnwidth,]{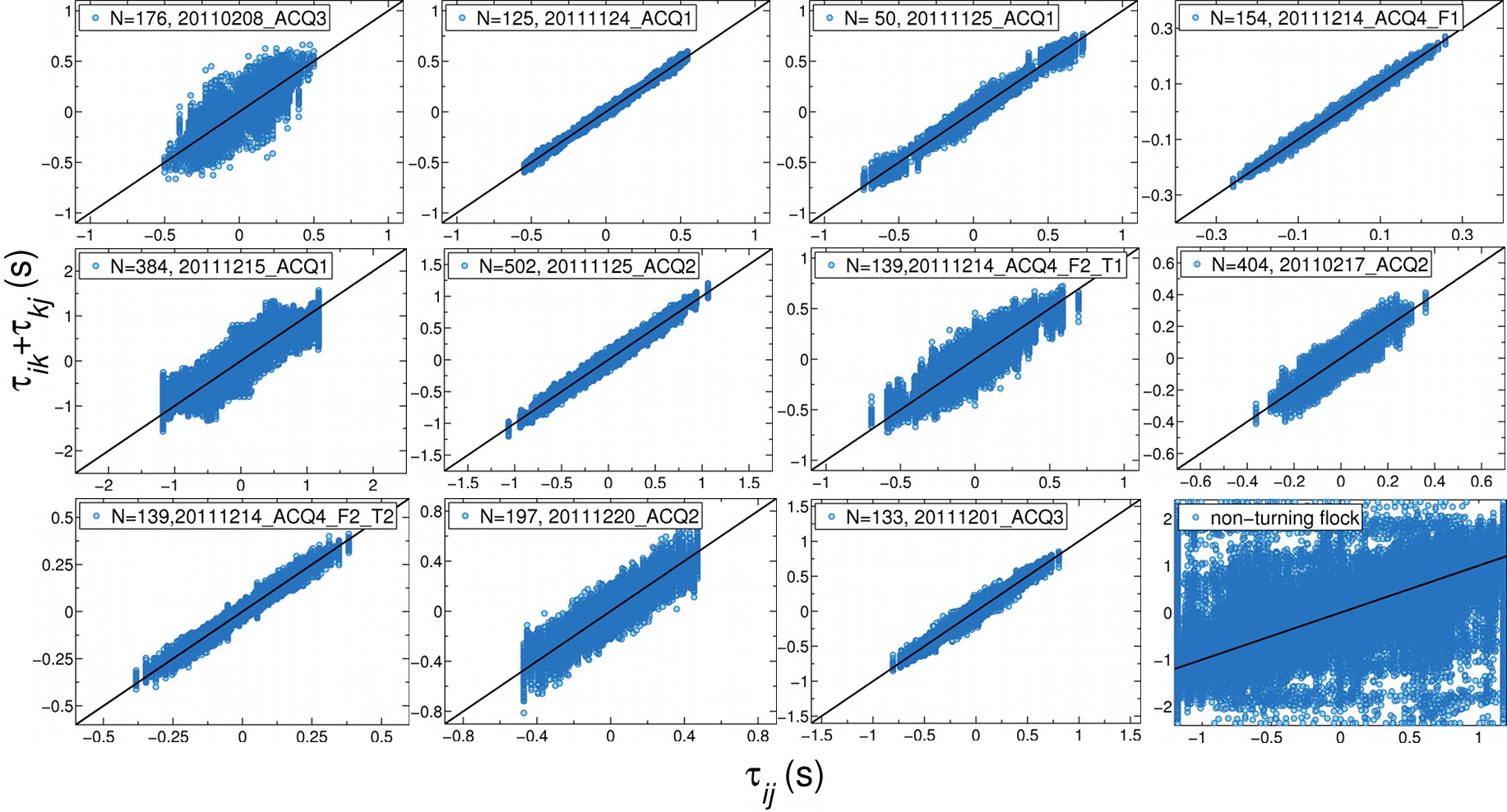}
 \caption{{\bf Check of time ordering relation (TOR).} We report the TOR test for several of our flocks and for one non-turning flock (lowest-right panel). Temporal consistency requires that $\tau_{ij}\sim\tau_{ik}+\tau_{kj}$, so to have the data scattered along the identity line with clear correlation.
 In the case of the non-turning flock, on the other hand, the delays are just random numbers, so no temporal consistency is found.}
\label{fig:tor}
\end{figure*}



\begin{table}[h]
\vskip 0.1 in
\begin{tabular}{l|c|c|c}
\hline 
\hline 
 {\sc Event  label} &\hspace{0.2cm}$N$\hspace{0.2cm} & \hspace{0.2cm}$\Phi$\hspace{0.2cm} & \hspace{0.1cm} $c_s$ (ms$^{-1}$) \\
 \hline
  \hline
20110208\_ACQ3 & 176 & 0.806 & 10.09 \\
 \hline
20111124\_ACQ1 & 125 & 0.959 & 21.32  \\
 \hline
20111125\_ACQ1 & 50 & 0.866 & 16.19 \\
 \hline
20111215\_ACQ1 & 384 & 0.801 & 11.37 \\
 \hline
20111125\_ACQ2 & 502 & 0.841 & 11.93\\
 \hline
20110217\_ACQ2 & 404 & 0.854 & 18.85 \\
 \hline
20111214\_ACQ4\_F1 & 154 & 0.940 & 19.23 \\
\hline
20111214\_ACQ4\_F2\_T1& 139 & 0.890 & 18.66 \\
\hline
20111214\_ACQ4\_F2\_T2 & 139 & 0.808 & 17.70 \\
\hline
20111220\_ACQ2 & 197 & 0.907 & 13.77 \\
\hline
20111201\_ACQ3\_F1 & 133 & 0.793 & 9.41 \\
\hline
20110211\_ACQ1 & 595 & 0.757 & 10.98 \\
\hline
\hline 
\end{tabular}
\caption{{\bf Polarization and speed.}
$N$ is the number of birds in the flock. The polarization is defined as,
$\Phi = ||(1/N) \sum_i {\bf v}_i/v_i ||$. The values of $\Phi$ reported here are on average smaller than those reported in previous investigations
\cite{ballerini+al_08a,ballerini+al_08b}. The reason for this is that previous data were obtained with cameras sampling at $10$Hz, 
whereas the present data are obtained at $170$Hz. At this sampling frequency experimental noise and wing flapping reduce the polarization.
This reduction, however, affects equally all flocks by simply rescaling $(1-\Phi)$, hence it does not change the correlation in Fig.\ref{fig:prediction}.
The speed of propagation of the information, $c_s$, is found by fitting the linear regime of the propagation curve, $x(t)$.
We note that by fitting the ranking curve to a power law for early-intermediate times, $r(t) = t^\alpha$, we find on average $\alpha=3.2$, thus justifying the 
statement that $x\sim r^{1/3}\sim t^{1.07}$ is a linear function.
\label{table:flocks}}
\end{table}

\setcounter{section}{13}  

\section*{Methods}
\label{methods}

{\bf Experiments}.
European starlings ({\it Sturnus vulgaris}) spend the winter in Rome, where they populate several roosting sites. Data were collected at the site of Piaz\-za dei Cinquecento, between November 2010 and December 2012. To acquire the video sequences we use the trifocal stereometric setup described in \cite{cavagna+al_08a}. Two cameras separated by a baseline distance $D_{12}=25$m are the stereometric pair. A third camera, placed at a shorter distance $D_{23}=2.5$m from the first one, allows us to exploit the trifocal constraint for solving the stereo correspondence (matching) problem  \cite{cavagna+al_08a}. We employ three high--speed cameras IDT-M5 with monochromatic CMOS sensor with resolution $2288\times1728$ pixels, shooting at $170$hz. Cameras store images on off-board memory using the Camera Link protocol. Lenses used are Schneider Xenoplan $28$mm $f/2.0$. Typical exposure parameters are: aperture between $f/2.8$ and $f/8$; exposure time between $700$ and $3500~\mathrm{ms}$. Intrinsic camera parameters are calibrated in the lab using a set of images of a known planar target. The recorded events have a time duration between $1.8$ and $12.9$ seconds. The data-set consists of 12 distinct flocking events, each one including one collective turn. If on the recorded sequence the flock performs more than one turn, the time lag is chopped and different turns are studied as independent events.


{\bf Tracking}.
Detection of individual birds on the images is carried out by using the same method as in \cite{cavagna+al_08a}. To assign stereoscopic links, i.e. to match birds across the three images, we use global optimization using a cost function based on the trifocal constraint~\cite{hartley2003book}. To perform temporal linking we first use a roto--scale--translation \cite{kabsch1976ac} to predict the position of each bird in the next frame. We then link birds from one frame to the next one in a redundant way, i.e. when in doubt we use multifurcation. Percolating the full set of temporal links though the entire temporal sequence, we construct all the possible $2d$ paths in the image space of each one of the three cameras. The three sets of $2d$ paths are then matched via a global assignment, based on a cost function which measures the number of  stereoscopic links between each triplet of $2d$ paths. Global optimization not only matches the right $2d$ paths, producing the correct $3d$ trajectories, but also eliminates the unphysical paths. To avoid exponential explosion of the number of paths, the temporal sequence is recursively divided into shorter time intervals over which the complexity of global optimization can be handled. All global optimizations are performed using linear programming \cite{cplex1994}. Full details of the tracking algorithm will be reported elsewhere.


{\bf Filtering}. 
Filtering of the time-discrete trajectories is necessary for two reasons: i) to reduce experimental noise; ii) to eliminate wing-flapping, whose frequency for starlings is $\omega_\mathrm{flap}=10$hz. By sampling at $170$hz we are fully exposed to the trajectories zig-zag (see inset in Fig.\ref{fig:methods}a), which would completely dominate acceleration. To cut this high frequency mode we a use $2$nd order lowpass digital Butterworth filter on the velocities, typically with a cutoff frequency $\omega_\mathrm{flap}/30$. Accelerations obtained in this way (see Fig.\ref{fig:methods}d) capture the low frequency directional change corresponding to the turn. Our final results are robust against changes of the cutoff frequency.


{\bf Turning delay}.
We define the turning delay $\tau_{ij}$ of bird $i$ with respect to bird $j$ as the time by which we have to shift the radial acceleration $a_j(t)$ with respect to $a_i(t)$ to maximally overlap them. More precisely, we define the following normalized correlation (or overlap) function,
\begin{equation}
 G_{ij}(\tau) = \frac{
                       \langle {\bf{a}}_{i}(t) \cdot {\bf{a}}_{j}(t-\tau) \rangle 
                     - \langle {\bf{a}}_{i}(t) \rangle \cdot \langle {\bf{a}}_{j}(t-\tau) \rangle }
    { \sigma_i \sigma_j } \ ,
    \nonumber
\end{equation}
where $\langle\cdot\rangle$ indicates a time average/integral and $\sigma_i$ is the fluctuation of ${\bf{a}}_i(t)$ during the turn, 
\begin{equation}
\sigma_i = \sqrt{\left(\langle{\bf{a}}_{i}(t)^2\rangle-\langle{\bf{a}}_{i}(t)\rangle^2\right)} \ . 
\end{equation}
Given this definition, the time shift $\tau_{ij}$ corresponds to the value of $\tau$ where $G_{ij}(\tau)$ reaches its maximum (see inset in Fig.\ref{fig:methods}d). $\tau_{ij}>0$ means that $j$ turns before $i$, and vice versa.
In absence of noise time ordering relation -TOR - requires that $\tau_{ij}=\tau_{ik}+\tau_{kj}$, for each triplet $i,j,k$. We check robustness of this relation with respect to noise in all our flocks and find a relatively small spread of the data along the identity line (Fig.~\ref{fig:tor}). We recall that $\tau_{ij}$ uses the full trajectory information of the two birds, which in turn is the product of the raw field data, of the tracking method and of the time-discrete data filtering.  Hence, by proving that the TOR violation is low, we give a rather strong proof of reliability of our entire experimental method. The quality of our TOR consistency test can be fully appreciated when we compare turning with non-turning flocks. If there is no turn $\tau_{ij}$ is just a random number, so temporal consistency is strongly violated and the TOR consistency plot really looks quite different (Fig.~\ref{fig:tor}, lowest-right panel).


\begin{figure*}[t!]
  \centering
  \includegraphics[width=1.8\columnwidth,]{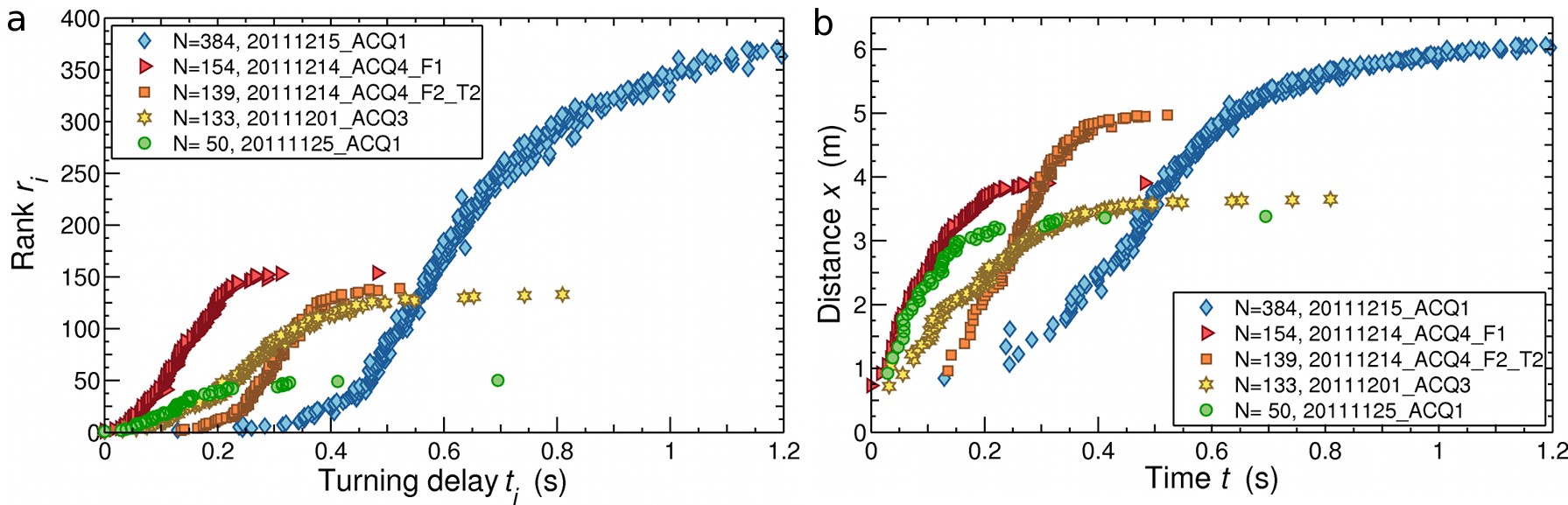}
 \caption{{\bf Ranking and propagation.} The ranking curve, $r(t)$ and the propagation curve, $x(t)= [r(t)/\rho]^{1/3}$, are reported for several turning flocks in our pool of data.}
\label{fig:allranking}
\end{figure*}

{\bf Ranking}.
In absence of noise TOR would be exactly satisfied and ranking would be trivial. However,  noise introduces some violations of TOR and we are in a similar case as sport ranking, where player $i$ may win over $k$, $k$ may win over $j$, but $i$ may lose to $j$, introducing some frustration. 
For every bird $i$, we say that $i$ `wins' over $j$ if
$\tau_{ij}<0$, in which case we set $J_{ij}=1$; conversely, $i$
`loses' to $j$ if $\tau_{ij}>0$, in which case we set $J_{ij}=-1$. We
then define the total score of $i$ as, $\phi_i = \sum_{j\neq i} J_{ij}$. 
Given that we are in a {\it round-robin} tournament, 
it makes sense to rank the birds simply according to the 
scores $\phi_i$ \cite{conner_00}. Thanks to the low violation of TOR, 
this score ranking already gives very small frustration (defined as the number of
cases in which $i$ ranks higher than $j$, but $i$ has lost to $j$). More refined
rankings can be obtained by using probabilistic methods \cite{conner_00}. 
In our case, these methods decrease only marginally the (already low)
frustration.

{\bf Absolute turning time}.
The absolute turning time $t_i$ for each bird $i$ is the delay with respect to the top bird in the ranking, i.e.\ the first to turn ($t_\mathrm{top} = 0, \ r_\mathrm{top}=1$). 
However, to reduce the statistical error on $t_i$ introduced by TOR violations we define $t_i$ using the mutual delay $\tau_{ij}$ with respect to {\it all} birds $j$ better ranked than $i$,
\begin{equation}
  t_i = \frac{1}{r_i-1} \sum_{r_j<r_i} (t_j+\tau_{ij}) \quad , \quad r_i>1  \ .  
 \label{vodka}
\end{equation}
Clearly, if there were no TOR violations, we would simply have $t_i=\tau_{i,\mathrm{top}}$. In the presence of noise, though, definition \eqref{vodka} is a more
reliable estimate of $t_i$. By plotting $r_i$ vs. $t_i$ for all birds in the flock, we obtain the ranking curve, $r(t)$, which is reported for several of our flocks in Fig.~\ref{fig:allranking}  together with the propagation curve, $x(t) = [r(t)/\rho]^{1/3}$.


\end{document}